\documentstyle[preprint,aps]{revtex}
\begin{document}
\draft
\def\thefootnote{\fnsymbol{footnote}}
%\parindent 30pt\textheight 8in\topmargin 0in\textwidth 6in
%\oddsidemargin .25in\evensidemargin 0in
%%%%%%%%%%%%%%%%%%%%%
% Defined in Rome
%%%%%%%%%%%%%%%%%%%%%
\def\lsim{\mathrel{\mathchoice {\vcenter{\offinterlineskip\halign{\hfil
$\displaystyle##$\hfil\cr<\cr\sim\cr}}}
{\vcenter{\offinterlineskip\halign{\hfil$\textstyle##$\hfil\cr
<\cr\sim\cr}}}
{\vcenter{\offinterlineskip\halign{\hfil$\scriptstyle##$\hfil\cr
<\cr\sim\cr}}}
{\vcenter{\offinterlineskip\halign{\hfil$\scriptscriptstyle##$\hfil\cr
<\cr\sim\cr}}}}}

\def\gsim{\mathrel{\mathchoice {\vcenter{\offinterlineskip\halign{\hfil
$\displaystyle##$\hfil\cr>\cr\sim\cr}}}
{\vcenter{\offinterlineskip\halign{\hfil$\textstyle##$\hfil\cr
>\cr\sim\cr}}}
{\vcenter{\offinterlineskip\halign{\hfil$\scriptstyle##$\hfil\cr
>\cr\sim\cr}}}
{\vcenter{\offinterlineskip\halign{\hfil$\scriptscriptstyle##$\hfil\cr
>\cr\sim\cr}}}}}

\narrowtext
\def\be{\begin{eqnarray}}
\def\ee{\end{eqnarray}}

%\begin{document}
%\noindent \hfill{Preliminary Draft(August, 1997)}
%\vskip 1.8cm

\title{\large \bf Primordial Abundances of $^6$Li, $^9$Be, and $^{11}$B
with\\ Neutrino Degeneracy and  Gravitational Constant Variation}

%\footnote{Talk presented at
%the Korea-Italian meeting on the Relativistic Astrophysics,
%Seoul and Suanbo, 1-5 Sept., 1997}

\vskip 1.0 cm
\author{Jong Bock Kim,$^{a,b}$ Joon Ha Kim,$^{a}$ and Hyun Kyu Lee$^{a}$}
\vskip 0.5 cm
\address{${}^a$ Department of Physics, Hanyang University,
Seoul 133-791, Korea}

\address{${}^b$ Department of Physics, Soonchunhyang University,
Asan, Choongnam 336-745, Korea} 
\maketitle
\vskip 0.8cm

\begin{abstract}
%\centerline{\bf Abstract}
\vskip 1cm
\noindent

 Recent measurements of deuterium abundances from QSO absorption spectra
 show
two conflicting numbers, which differ by an order of magnitude.
 Allowing the
 neutrino degeneracy together with gravitational
constant variation at the epoch of primordial nucleosynthesis,
the implication
of these observations on
the $^6$Li, $^9$Be and $^{11}$B abundances
will be discussed.  Within the permitted ranges
consistent with  D, $^4$He and $^7$Li, we observe the strong
$\eta$ dependence of  $^{11}$B as in SBBN but no significant
dependence of  $^6$Li and $^9$Be on the baryon number density is observed.
 The predictions of $^6$Li and  $^9$Be for low and high deuterium differ by
an order of magnitude
and predictions for $^{11}$B differ by several orders of magnitude at most
and vary with $\eta$.

\end{abstract}
\pacs{PACS numbers: 26.35.+c, 26.45.+h, 97.10.Cv, 97.10.Tk, 98.80.Ft}

\newpage
%\section{Introduction}
     The primordial abundances of the light nuclei
has been considered as the important observables, which probe the physical
environments of  the early Universe in the frame work of  hot Big-Bang
cosmology \cite{sm}.  One of the successes of the standard  Big-Bang
nucleosynthesis(SBBN)\cite{bbnr} is the determination of the baryon number
 density, $n_B$, which take part in the process of
nucleosynthesis. Conventionally the baryon number density is
expressed as the ratio to the photon number density, $\eta= n_{B}/n_{\gamma}$
\cite{ytsso}  It measures not only the degree of baryon number asymmetry
\cite{baryon} but also  determines baryonic
 energy density of Universe, which
puts the possible limit of the  baryonic dark matter  in the
Universe.
The ratio of the baryonic energy density($\rho_B$) to the
critical
energy density($\rho_{c}$),  $\Omega_{B}(=\rho_{B}/\rho_{c})$,
 can be written in terms of
$\eta$ as
$\Omega_B h_0^2 \sim  0.015 \, \eta_{10}$, where
the Hubble parameter $ H_{0}=50h_{0}km^{-1}Mpc^{-1}$
and $\eta_{10} = \eta \times 10^{-10}$.

Recent measurements of deuterium
abundances from QSO absorption spectra \cite{Rug,Tyt}  show
two conflicting numbers which differ by an order of magnitude.  Since the
deuterium abundance is known to be very sensitive on the baryon number 
density in contrast to the $^4$He abundance in
the range of interest, the implication of these two measurements
is  very important in determining the baryon number density.  It is also
well known that the heavier elements like $^9$Be and
$^{11}$B are sensitive on the baryon number density, change of abundances
by several orders of magnitude for $\eta = 1 - 10$.
Recently, Nollet et al.\cite{nls}  discussed  that $^6$Li abundance
is  also very sensitive on the baryon
number density, $\eta$, within the uncertainties of the nuclear reaction rates
 relevant to $^6$Li production. Hence these heavier elements are considered
to be very useful
in discussing the possible variations from SBBN, since  some of the
variations allow different
ranges of baryon number density. For example, there has
been some works on the  $^9$Be and $^{11}$B with
inhomogeneous baryon number density possibly due to the first order QCD phase
transition\cite{inhomo}.  In this work, we will consider the case with
neutrino degeneracy\cite{Kang}
together with the gravitational constant variation\cite{Kim} to investigate
the implications of
 two distinct measurements of deuterium abundances on the primordial
abundances of $^6$Li, $^9$Be and $^{11}$B.

In the radiation dominated era of early Universe, the evolution
is described by the expansion rate $H$ which is defined as
$H^{2} = \frac{8\pi}{3} G \rho_{rad}$, where
$G$ is the gravitational constant and
$\rho_{rad}$ is the energy density of the light particles.
In SBBN, photons, electrons and
three light neutrinos are only considered. At BBN epoch , $T \sim 1MeV$,
muon and tau ($m_{\mu}$ and $m_{\tau}\gg 1 MeV$)  have been already decayed
away and their chemical potentials can be put to zero. Assuming the charge
neutrality of the Universe the electron chemical potential can be also
ignored, since $\frac{\mu_{e}}{T}$ should
be of order $\cal{O}(\eta)$.  However, there is presently no
convincing theoretical or observational constraints on the neutrino
degeneracies at the
epoch of primordial nucleosynthesis\cite{Mal}
 and we take it as parameters
of the early environment of the Universe.
Neutrino degeneracy
increases $\rho_{rad}$ and therefore speeds up the expansion of the
Universe. One can easily see also that the deviation of
gravitational constant $G$ at the epoch of  nucleosynthesis from the present
 value $G_{0}$
can also influence the expansion rate of the Universe and hence the standard
 BBN\cite{yssr}.
Since the effects of
degeneracies of muon- and tau-neutrinos are only on the energy density,
it can be effectively implemented as the variation of gravitational constant.
The degenerate electron-neutrino not only increase $\rho_{rad}$ as muon- and
tau-neutrinos but also it influences the weak interaction rates where
electrons are involved. Hence
the role of  degenerate electron neutrinos cannot be simply absorbed in
$G$ variation. In this work, we consider only
 electron-neutrino degeneracy  together with the variation of gravitational
constant, $G$,
possibly due to the degenerate muon- and/or
  tau-neutrino  or more fundamental origins\cite{jbd}.
 The three light neutrinos are
 considered to be effectively massless during
the nucleosynthesis and no neutrino oscillation is assumed.

For deuterium abundances, we take both high and low values from QSO
 observations.
The low abundances of deuterium, [D/H]=$(1.7-3.5)\times 10^{-5}$ which
is close to the ISM value, has been estimated by Tytler et al. \cite{Tyt},
while higher
abundance,   [D/H]=$(1.5-2.3)\times 10^{-4}$ is obtained by Rugers et al.
\cite{Rug}. We take
$0.226\leq Y_p \leq 0.242$ for $^4$He and 0.7$\leq$ [$^7$Li/H]$\times 10^{10}
\leq 3.8$
for $^7$Li\cite{Oliv}.
We obtain  the allowed ranges for $\eta_{10}$, $G$, and $\xi_e$
which are consistent with the above observed abundances
\footnote{The variation
 of gravitational constant is chosen to be $G/G_0 = 0.1 - 6$ for numerical
convenience, which  is enough to see the dependences  on the relevant
parameters of the abundances.}.
The result\cite{jjl97} is
summarized in $\xi_e - G$ plot of Fig. 1.  It shows
a rather wide range of possible  $\xi_e$ and $G$ for given $\eta_{10}$
(numbers
in the shaded boxes for low deuterium abundances and empty boxes for higher
abundances). And It also demonstrates clearly how these two
distinct observations of  D/H abundance predict the different ranges
of $\eta$: smaller(larger) value of
$\eta_{10}$ for high(low) deuterium abundance.
One can see that upper bound  on the baryon
energy density with low deuterium abundance,  $\Omega_Bh_0^2 \leq 0.15
(\eta_{10}=10)$, is larger than
that of standard BBN.

To see the implication of these two
different ranges
 of $\eta$,
we calculate  the primordial abundances of
$^6$Li, $^9$Be and $^{11}$B
using these permitted sets  of parameters obtained above from the observed
abundances of D,
$^4$He and $^7$Li.  In Fig. 2, the calculations of $^9$Be abundance
are shown. In SBBN, it decreases monotonically as $\eta$ increasing and
is quite sensitive on $\eta$, which is the only parameter of SBBN. However, if
we allow neutrino degeneracy and gravitational constant variation, we found
that the
predictions are not so sensitive on $\eta$. It is  almost independent of
$\eta$ both for high and low deuterium cases. For low deuterium case, the
abundance is similar to that of SBBN with $\eta \sim 6$
and for high deuterium case the
 abundance
is similar to that of SBBN with $\eta \sim 2$, whatever the value of $\eta$
is used in the permitted ranges of $\eta, G/G_0$ and $\xi_e$.  This means
that $^9$Be does not constrain the $\eta$ any further as far as the parameter
 sets are  consistent with D,
$^4$He, and $^7$Li.   However one can clearly see the differences in the
abundance predictions with the low and high deuterium cases, which differ
 by an order of magnitude.   Fig. 3 shows SBBN predictions of   $^{11}$B
abundance which  are changing rapidly with $\eta$. Allowing
neutrino degeneracy and gravitational constant variation, the predictions
of low and high deuterium cases are found to be quite different. With high
deuterium, the sensitivity on $\eta$ get reduced as in $^9$Be as far as the
 relevant parameters are confined to be consistent with abundances of  D,
$^4$He, and $^7$Li. For low deuterium case, however,  the sensitivity on
$\eta$ of $^{11}$B abundance is still clear as in SBBN
within the permitted
ranges of $\eta$.
It is interesting to note that for $\eta_{10}
\sim 2$ the predictions with low and high deuterium becomes  similar.
    We include the reaction rates proposed by Nollet
et al. into nuclear reaction chains\cite{Kim} to calculate the possible
abundance of $^6$Li, which is monotonically
decreasing by an order of magnitude in the range of interest
 as $\eta$ increases, as shown in Fig. 4.  But the dependence on $\eta$ which
 is constrained to vary together with  $G/G_0$ and $\xi_e$ is not the same
as in SBBN.  The abundance for low deuterium remains roughly fixed at the
value of SBBN with $\eta \sim 6$ and at the SBBN value with  $\eta \sim 2$
for high deuterium.  They are differ by an order of magnitude but they show
no $\eta$ dependence as for $^9$Be.

In summary, we calculate
the abundances of $^6$Li, $^9$Be and $^{11}$B, using the permitted ranges
 of $\eta$ consistent with   D, $^4$He and $^7$Li
by allowing the
neutrino degeneracy and gravitational constant variation.
We investigate the implication of two different QSO measurements of
deuterium abundances
and the  possible $\eta$ dependence on those heavier elements.  Within the
permitted ranges of $\eta$, $^6$Li and $^9$Be do not show any significant
$\eta$ dependence. Within the allowed ranges obtained  for high deuterium
abundance,  $^{11}$B show essentially no dependence on $\eta$. However,
if we use the permitted
 ranges consistent  to low deuterium abundance, $^{11}$B shows clearly
the rapid dependence on $\eta$ as much as in SBBN.
It is found that the
predictions of the abundances of $^6$Li and  $^9$Be with low and
high deuterium abundance differ by an order of magnitude
and predictions for $^{11}$B differ by several orders of magnitude at most.
Therefore, the measurement of the heavier elements like  $^6$Li, $^9$Be and
 $^{11}$B will provide us a valuable test of the neutrino
degeneracy and gravitational constant variation proposed in this work and
also be useful in resolving the issue of the two different QSO
measurements of deuterium abundance .

\vskip 0.5cm
\centerline{\large\bf Acknowledgements}

This work is supported in part by Ministry
of Education(BSRI 97-2441) and by Korea Sceince and Engineering Foundation
(94-0702-04-01-3 and Center for Theoretical Physics at Seoul National
University).

\newpage
{\large\bf Figure Captions }
\vskip 0.5cm

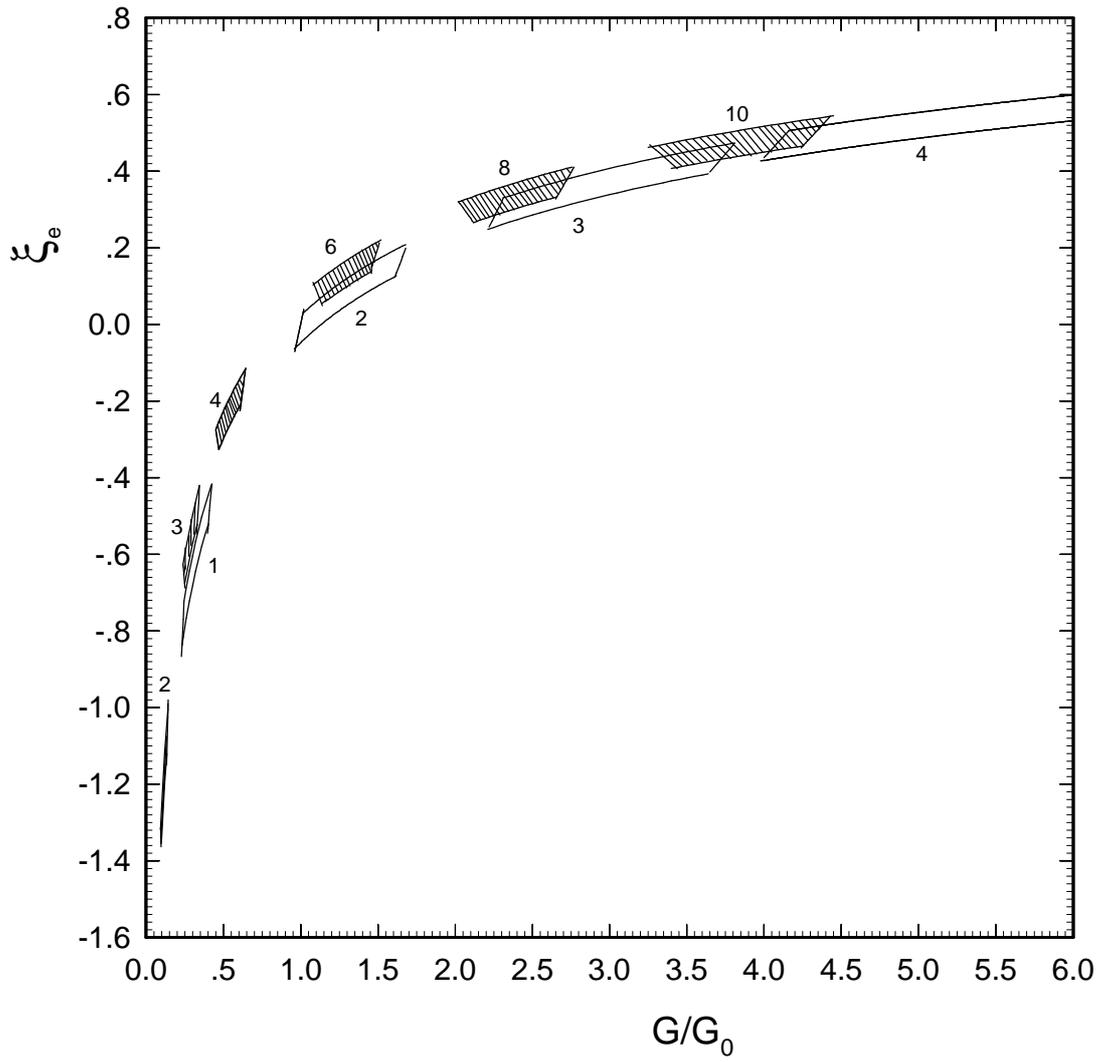
\noindent Fig. 1. Allowed ranges of $\xi_{e}$ and $G/G_{0}$\cite{jjl97}.
Shaded boxes
 for low deuterium  abundance and empty boxes for high deuterium abundance.

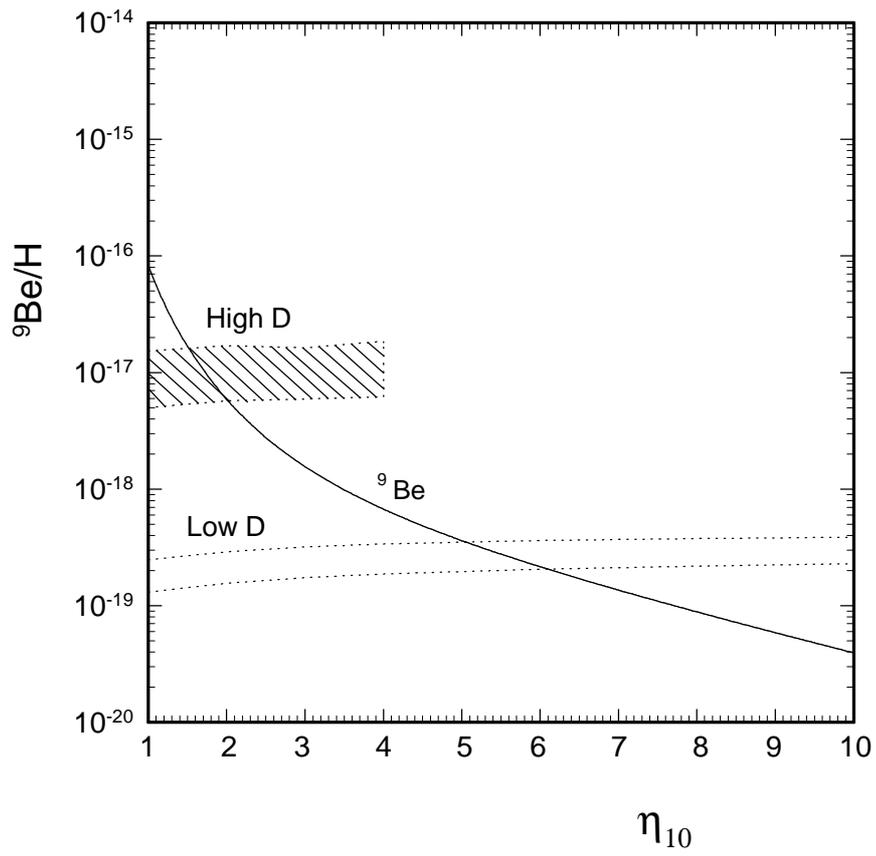
\noindent Fig. 2. The predictions of $^9$Be abundances;
solid line for SBBN and  shaded(empty) band for high(low) deuterium abundance.

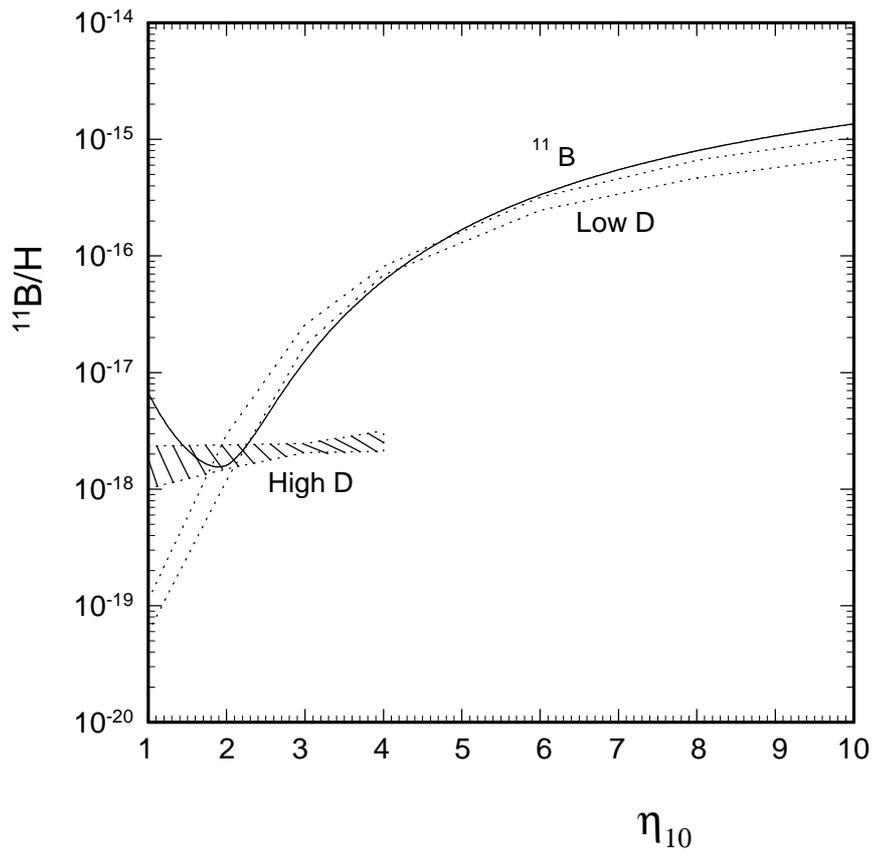
\noindent Fig. 3.  The predictions of $^{11}$B abundances;
solid line for SBBN and  shaded(empty) band for high(low) deuterium abundance.

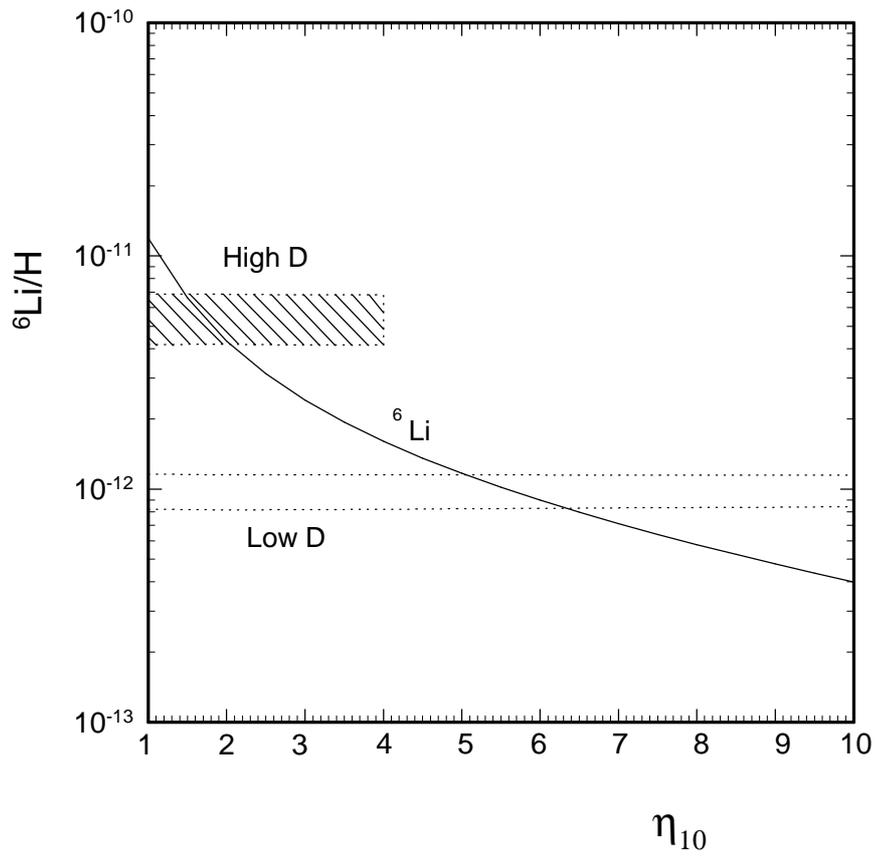
\noindent Fig. 4.  The predictions of $^6$Li abundances;
solid line for SBBN and  shaded(empty) band for high(low) deuterium abundance.

%%%%%%%%%%%%%%%%
%%%%%%%%%%%%%%%%

\end{document}